\documentclass{aa}
\usepackage{graphics}

\newsavebox{\junk}

\newcommand{\Ha}{H$\alpha${}} 
\newcommand{\Hb}{H$\beta${}}

\begin{document}

\thesaurus{06         
              (08.22.2;   
               08.15.1;   
               08.09.2;   
               03.20.8)} 

\title{Oscillation mode identifications and models for the $\delta$~Scuti
star FG~Virginis}

\author{M. Viskum\inst{1}, 
        H. Kjeldsen\inst{1,2}, 
        T.R. Bedding\inst{3},
        T.H. Dall\inst{1}, 
        I.K. Baldry\inst{3}, 
        H. Bruntt\inst{1}  and 
        S. Frandsen\inst{1}
       }

\offprints{M. Viskum}

\institute{Institute of Physics and Astronomy, Aarhus University, DK-8000 
           Aarhus C, Denmark, \and Theoretical Astrophysics Center, 
           Danmarks Grundforskningsfond, Aarhus University, DK-8000 Aarhus C,
           Denmark \and Chatterton Astronomy Department, School of Physics, 
           University of Sydney, 2006, Australia}

\date{Received ; accepted date}
\authorrunning{M. Viskum et al.}
\titlerunning{The $\delta$~Scuti star FG~Vir}

\maketitle

\begin{abstract}
We present new spectroscopic and photometric time series observations of
the $\delta$ Scuti star FG~Vir.  We detect the oscillations via changes in
the equivalent widths of hydrogen and metal absorption lines.  {}From the
ratios between spectroscopic and photometric amplitudes, we assign $\ell$
values to the eight strongest oscillation modes.  In particular, we
identify two radial modes ($\ell =0$) and find that the main pulsation mode
(147~$\mu$Hz) has $\ell =1$.  One of the radial modes (at 140$\mu$Hz) is
the fundamental, implying that two modes with lower frequencies are {\it
g}-modes.  For the radial modes, we compare frequencies with those
calculated from a scaled $\delta$~Scuti star model and derive a density
$0.1645\pm 0.0005\,\rho_{\odot}$.  We then obtain a distance of $84\pm
3$\,pc, in excellent agreement with the Hipparcos value.  Finally, we
suggest that a 3.5-day variability in all observables (equivalent widths
and intensity) is caused by stellar rotation.

\keywords{stars: $\delta$~Scuti  --  
          stars: oscillations  -- 
          stars: individual:HD106384=FG~Vir -- 
          techniques: spectroscopic
         } 
\end{abstract}

\section{Introduction} \label{intro}

Oscillations in multi-periodic variables such as $\delta$ Scuti, roAp and
$\beta$ Cephei stars have been observed extensively during the past 20
years.  But even with high-quality data, it is still extremely difficult to
identify which modes are being detected.  Kjeldsen et al.\ (\cite{kbvf95})
used a new technique to detect solar-like oscillations in the bright G
sub-giant $\eta$ Boo through their effect on the equivalent widths of the
Balmer lines.  A subsequent discussion by Bedding et al.\ (\cite{bkrb96})
of the sensitivity of different observables to modes with different degree
$\ell$ suggested that one can determine the $\ell$-value of a given mode by
combining measurements of absorption-line equivalent widths with
simultaneous photometric observations.  To test this idea, we chose the
bright and well-studied $\delta$~Scuti star FG~Vir.

FG Vir (HD 106384; $V=6.57$) is a multi-periodic $\delta$~Scuti star.  It
has a main pulsation period close to 1.9 hours and shows a fairly complex
oscillation spectrum.  This star has been studied extensively during the
last few years, resulting in the detection of at least 24 well-determined
frequencies between 100 and 400~$\mu$Hz, with amplitudes from 0.8 to 22
milli-magnitudes (mmag; Breger et al.\ \cite{bhn95}, \cite{br98}).  Because
of its slow rotation (which reduces the complicating effects of rotational
splitting) and the large number of detected frequencies (some of which are
probably {\it g} modes), FG Vir is one of the most promising candidates for
performing asteroseismology on a $\delta$ Scuti variable.  Observations and
models of this star have been presented by Dawson et al.\ (\cite{dbl95}),
Breger et al.\ (\cite{bhn95}) and Guzik \& Bradley (\cite{gb95}).

By choosing a star like FG Vir we have the advantage of knowing the
frequencies in advance. We are therefore able to determine the oscillation
amplitudes and phases with high precision.  The aim of this paper is to
identify the $\ell$ values of the observed modes and to
compare the oscillation frequencies with a pulsation model.
A preliminary analysis of the observations presented in this paper was
given by Viskum~et~al.~(\cite{vdk97}), while results on radial velocity
measurements were given by Viskum et al.\ (\cite{vb97}).

\section{Observations and basic data reduction} \label{obs}

\subsection{Spectroscopy}

We obtained intermediate-resolution spectra of FG Vir using the DFOSC
spectrograph mounted on the Danish 1.54-m at La Silla, Chile and the
coud\'e spectrograph (B grating) on the 74-inch Telescope at Mt.\ Stromlo,
Australia, during a period of more than two weeks in May 1996. A journal of
the observations is given in Table~\ref{log1}.  In total we obtained 652
spectra of FG~Vir.

The DFOSC spectrograph was used in the echelle mode, covering a wavelength
region from 4600\AA{} to 8000\AA{} at a resolution of $R\sim 4300$.  Six
echelle orders were recorded, each covering about 600\AA{} and with an overlap
between adjacent orders of about 370~\AA.  A S/N ratio of 250--360 
was reached in the DFOSC data with a typical exposure time of 460\,s.
The S/N ratio was determined by measuring the standard deviation of the
intensity of the normalized spectrum in two spectral windows free of
absorption lines.  

The Mt.\ Stromlo data consisted of single-order spectra, with a wavelength
coverage of 6000\AA{} to 7000\AA{}. The dispersion was 0.49\AA{}/pixel,
with a resolution of about 1.5\AA{} which was set by the slit width of 2
arcsec.  The typical exposure time was 400\,s.

\begin{table}
 \caption[ ]{Journal of spectroscopic observations of FG~Vir} \label{log1}
 \begin{flushleft}
  \begin{tabular}{llccc}  \hline
    \qquad Date&  Site    & HJD (start)& Length& No. of  \\
    \qquad UT &          & 2450000 +  & hours & spectra  \\
    \hline
   1996 May  2 & Stromlo  & 205.931    &4.0    & 26 \\
   1996 May  3 & Stromlo  & 206.834    &4.2    & 31 \\
   1996 May  6 & Stromlo  & 209.867    &7.7    & 62 \\
   1996 May  7 & Stromlo  & 210.838    &5.9    & 47 \\
   1996 May  8 & La Silla & 212.484    &6.2    & 34  \\
   1996 May  9 & La Silla & 213.486    &6.3    & 45  \\
   1996 May 10 & La Silla & 214.496    &6.1    & 43  \\
   1996 May 11 & La Silla & 215.496    &6.0    & 50  \\
   1996 May 13 & La Silla & 216.550    &4.7    & 27  \\
   1996 May 13 & La Silla & 217.492    &6.0    & 41  \\
   1996 May 15 & La Silla & 218.525    &5.1    & 40  \\
   1996 May 15 & Stromlo  & 218.824    &6.2    & 69 \\
   1996 May 15 & La Silla & 219.468    &5.1    & 35  \\
   1996 May 17 & La Silla & 220.541    &3.3    & 20  \\
   1996 May 17 & La Silla & 221.477    &6.2    & 43  \\
   1996 May 18 & La Silla & 222.498    &5.5    & 39  \\
   \hline
  \end{tabular}
 \end{flushleft}
\end{table}

We used the IRAF package for bias subtraction, flat-field correction,
extraction of the 1D spectra and subtraction of background scattered light.
We found the Mt.\ Stromlo CCD to have a non-linearity of 7\% from low-light
levels to saturation, which we corrected in pre-processing.  The DFOSC CCD
camera was measured to be linear within 0.5\% and no correction for
non-linearities was performed.  

The DFOSC spectra suffered from sky contamination and edge vignetting,
which meant that the extracted spectra tapered from the centre to the edge.
We therefore normalized the continuum by dividing each order by a curve
containing the shape of the flat field.  The Mt.\ Stromlo spectra did not
suffer for this kind of edge vignetting, so a third order polynomial fit
was used for the continuum normalization.

\subsection{Photometry}

We obtained Str\"omgren $uvby$-$\beta$ photoelectric photometry with the
six-channel photometer mounted at the Danish 0.5-m reflector (SAT) at La
Silla, Chile over 22 nights in 1996.  Observations of FG~Vir were
interspersed with observations of the star HD~105912 (spectral type F5\,V).
This was used as a comparison star by Breger~et~al.~(\cite{bhn95}), who
found no variability.  In Table~\ref{log2} we present a journal of the
observations.  The precision per data point is about 3~mmag.  During three
of the nights, only the \Hb\ index was obtained due to non-photometric
weather conditions.  A total of 768 useful data points (99 hours) were
obtained.

To produce light curves from the raw data, we used a reduction program
developed for the SAT telescope. The reduction included airmass correction
using standard extinction coefficients and transformation to the standard
instrumental system.  Slowly varying effects were then removed by
subtracting second-order polynomials for both FG Vir and the comparison
star.  Variations in FG Vir not due to oscillations were removed by
subtracting the comparison star, resulting in the light curves whose
analysis is described in Sec.~\ref{analysis}.

\begin{table}
 \caption[ ]{Journal of photometric observations of FG~Vir} \label{log2}
 \begin{flushleft}
  \begin{tabular}{lccccc}  \hline
   \qquad Date   & HJD (start) & Filter   & Length & No. of  \\
   \qquad UT     & 2450000 +   &          &hours& data pts  \\
   \hline
   1996 Apr 28 &202.469 & uvby-$\beta$  &7.1      & 62 \\
   1996 May  1 &205.470 & uvby-$\beta$  &6.9      &69  \\
   1996 May  2 &206.472 & uvby-$\beta$  &6.8      & 46  \\
   1996 May  3 &207.465 & uvby-$\beta$  &7.2      & 48  \\
   1996 May  4 &208.490 &uvby-$\beta$&6.4      & 42 \\
   1996 May  5 &209.469 &$H{\beta}$&7.0  & 83 \\
   1996 May  7 &210.511 &$H{\beta}$&5.2  & 51 \\
   1996 May  8 &212.464 &uvby-$\beta$&6.9      & 46  \\
   1996 May  9 &213.465 &uvby-$\beta$&3.2      & 22 \\
   1996 May 10 &214.457 &uvby-$\beta$&7.3      & 48  \\
   1996 May 11 &215.458 &uvby-$\beta$&7.4      & 49 \\
   1996 May 13 &217.462 &uvby-$\beta$&7.2      & 45 \\
   1996 May 14 &218.486 &uvby-$\beta$&6.4      & 42 \\
   1996 May 15 &219.470 &$H{\beta}$& 4.2 & 51 \\
   1996 May 17 &221.461 &uvby-$\beta$&6.4      & 43 \\
   1996 May 18 &222.475 &uvby-$\beta$&3.1      & 21 \\
   \hline
  \end{tabular}
 \end{flushleft}
\end{table}

\section{Determination of equivalent widths}

Stellar acoustic oscillations ($p$~modes) can be measured via their effect
on the equivalent widths (EWs) of the Balmer lines, which are sensitive to
temperature.  Metal lines are also sensitive to temperature changes
(Bedding et al., \cite{bkrb96}), with the Fe\,{\sc i} lines being
particularly sensitive but in the opposite sense to the Balmer lines.  We
measured the equivalent widths of the metal lines by fitting profiles, but
this technique was found unsuitable for the much stronger Balmer lines, for
which we used a different method.  Both techniques are described in the
next sections.

\subsection{Equivalent-width variations of the Balmer lines}

As described by Bedding \& Kjeldsen (\cite{bk98}), a robust way of
estimating the EW of broad spectral lines involves a method analogous to
Str\"omgren H${\beta}$ photometry.  We have attempted several other
methods, such as fitting directly to the line profile, without reaching the
same internal precision.

Our adopted method involved calculating the flux (number of counts) in
three software filters, one centred on the line ($L$) and the others on the
continuum on the red ($C_R$) and the blue ($C_B$) sides of the line.  For
each spectrum, we first adjusted its slope so that the fluxes were equal in
the continuum filters: $C_R = C_B = C$.  We then calculated the quantity
\begin{equation}
 W = {(C-L)\over C}.
\end{equation}
This calculation was repeated for a large number of positions around the
line centre by moving the three filters together in wavelength by small
amounts to find the position which maximized the value of $W$.  The whole
procedure was repeated for different filter widths, in a way analogous to
aperture radii selection in CCD photometry, and the results combined to
give an estimate of the equivalent width (see Bedding \& Kjeldsen
\cite{bk98} for more details).

\begin{figure}
  \resizebox{8cm}{!}{\includegraphics[-40,0][264,380]{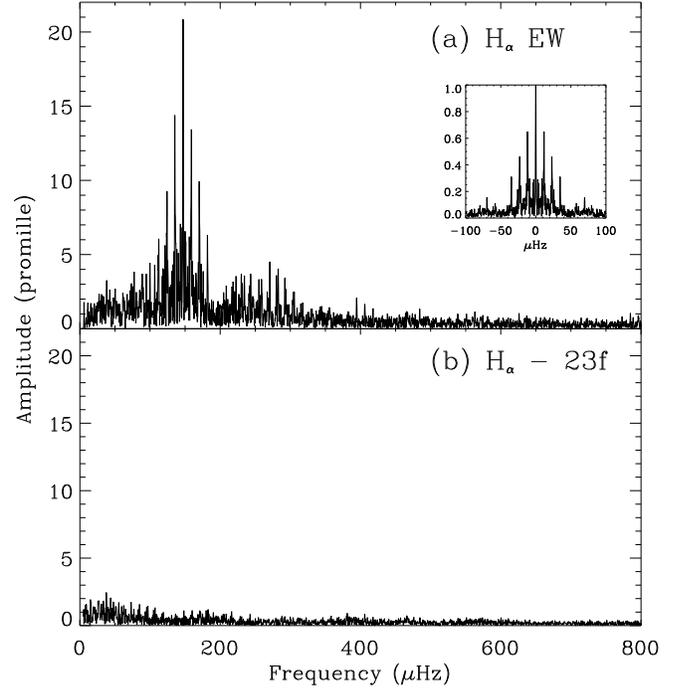}}
  \hspace{1cm} \caption[]{\label{ha}(a) Amplitude spectrum of the
  oscillations in FG~Vir measured through the EW of the \Ha\ line of the
  total data set (Mt.\ Stromlo and La Silla). The amplitude is measured in
  promille (parts per thousand).  The inset shows the amplitude spectrum of
  the window function.  (b)~Residual spectrum after subtraction of 23
  known oscillation frequencies. }
\end{figure}

Our La Silla spectra cover both the \Ha\ and \Hb\ lines, while only \Ha\ is
present in the Mt.\ Stromlo data.  In Fig.~\ref{ha}a we present the
amplitude spectrum of the \Ha\ EW for the combined data set (Mt.\ Stromlo
and La Silla), where the amplitude is measured in promille (parts per
thousand).  All amplitude spectra in this paper were calculated using a
weighted least-squares sine-wave fit to the time series (see Frandsen et
al.~\cite{kapboo}, \cite{6134.camp}).  The pulsation signal is clearly seen
in the range 100--400$\mu$Hz.  Fig.~\ref{ha}b shows the residual spectrum
obtained after subtracting 23 previously published frequencies (as
described in Sec.~\ref{analysis}).  The mean noise level in the residual
spectrum at high frequencies (600--800$\mu$Hz) is 0.18 promille, while it
is 0.35 promille in the frequency range where the oscillation signal is
found (100~--~400~$\mu$Hz).  The S/N of the dominant peak at 147~$\mu$Hz is
about 58.

\begin{figure}
  \resizebox{\hsize}{!}{\includegraphics[20,0][277,202]{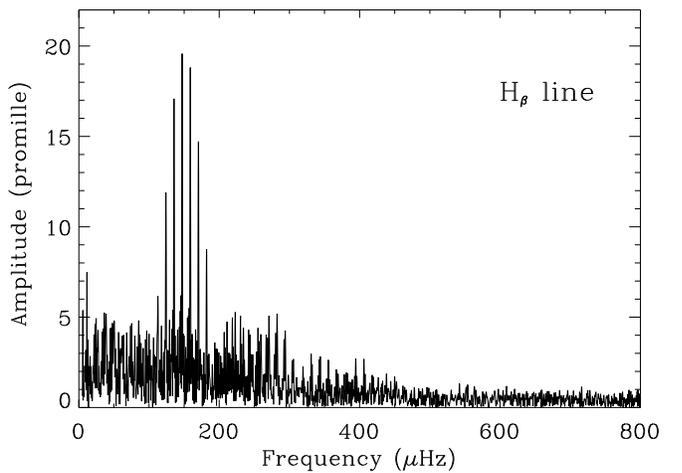}}
  \caption[]{Amplitude spectrum of the oscillations in FG~Vir measured
    through the EW of the \Hb\ line (La Silla data only).}
  \label{hb}
\end{figure}

Fig.~\ref{hb} shows the amplitude spectrum of the EW measurements of the
\Hb\ line (La Silla only).  The oscillations are again obvious, but the
noise is more than a factor two higher than for \Ha, due in part to bad CCD
columns that could not be removed by bias subtraction or flat fielding.

\subsection{Equivalent-width variations of metal lines}

The spectra from La Silla consist of six echelle orders containing many
metal lines.  The line identifications, kindly provided by F.~Kupka and
M.~Gelbmann, were made using the program described by Gelbmann
(\cite{gelb95}) and Gelbmann et al.\ (\cite{gelb97}).  We measured the EWs
by fitting Gaussians to the line profiles, with the surrounding continua
being fitted using low-order polynomials.  The fitting was done iteratively
using the Levenberg-Marquard method (Press et al.\ \cite{press}) to
estimate the line position, line depth and continuum level.  A first
estimate was made for these parameters over a narrow wavelength interval,
and an improved continuum was calculated over a broader range of the
spectrum. In addition to the EW, this also yielded other parameters to be 
used for decorrelation.

The time series of each line was decorrelated against external parameters.
A detailed discussion of our decorrelation procedure can be found in
Frandsen et al.\ (\cite{6134.camp}).  We used parameters that were derived
either from the spectra or from the CCD frames and were known {\em not\/}
to contain any information on the EWs or the pulsations.  Any correlation
between the measured EWs and these parameters must then have been due to a
systematic effect and was corrected by subtracting a simple linear fit.
The procedure was repeated until no significant correlations were present.
This procedure increased the S/N significantly: for example, the
improvement was a factor of 3.3 for the sum of all 20 lines described
below.

{}From the EW measurements we detected the strongest pulsation period of FG
Vir in all Fe\,{\sc i}, Ca\,{\sc i} and Mg\,{\sc i} lines, with amplitudes
from 7 and 22 promille.  Noise levels in the amplitude spectra ranged from
about 0.9 promille up to more than 3 promille for the weakest or highly
blended lines.

We also examined lines of ionized elements, which are potentially useful
for calibration.  The reason is that in FG Vir, these lines are near their
maximum strength as a function of temperature (Gray~\cite{gray}) and thus
are expected to be stable with respect to variations in the temperature
induced by the stellar pulsation.  We indeed found these lines to have very
low amplitudes in EW, with some of the Si\,{\sc ii} lines having no
detectable variability.  For example, the Si\,{\sc ii} line at 6372\AA{}
was stable in EW at the level of 2.7 promille (3$\sigma$).  We used the EW
of this line as a decorrelation parameter for the Fe\,{\sc i} lines.

\begin{table}
 \caption[]{\label{mtable} Amplitude of the strongest oscillation mode in
   FG~Vir as measured in EW for twenty strong Fe\,{\sc i} lines.  }

 \begin{tabular}{cccrc} \hline
$\lambda$         & Line  & depth & A$_{147}$& $\sigma$(A$_{147}$) \\ 
  (\AA)	          &  ID   & (\%)  & \multicolumn{2}{c}{(promille)} \\
\hline	 
  4958		  &  1-14 & 23 	  & 13.4 & 1.1  \\
   '' 		  &  2-26 & 24 	  & 13.0 & 1.5  \\
  4983		  &  2-25 & 14 	  & 18.5 & 1.2  \\
  5003		  &  2-23 & 15 	  & 15.1 & 1.7  \\
  5041		  &  1-19 & 15 	  & 13.1 & 1.2  \\ 
   '' 		  &  2-21 & 16 	  & 16.1 & 2.1  \\
  5227		  &  2-13 & 25 	  & 12.7 & 0.8  \\
  5270		  &  2-28 & 21 	  & 12.4 & 0.9  \\
  5329		  &  2-10 & 20 	  & 14.1 & 0.9  \\
  5341		  &  3-25 & 20 	  & 13.4 & 0.9  \\
  5370		  &  2-08 & 13 	  & 16.3 & 1.6  \\
  5404		  &  3-22 & 10 	  & 20.2 & 1.8  \\
  5406		  &  3-21 & 11 	  & 21.2 & 2.0  \\
  5424		  &  3-19 & 11 	  &  9.6 & 2.0  \\
  5430		  &  3-18 & 12 	  & 17.7 & 2.3  \\
  5456		  &  2-04 & 11 	  &  6.8 & 3.0  \\
   '' 		  &  3-15 & 13 	  &  9.0 & 2.0  \\
  5477\rlap{$^*$} &  2-03 & 11 	  & 17.8 & 2.2  \\
  5616		  &  3-05 & 11 	  & 15.1 & 1.7  \\
  5659		  &  3-03 & 10 	  & 22.0 & 1.6  \\
  \hline
 \end{tabular}

$^{*}$this line is blended with a strong Ni\,{\sc i} line.
\end{table}

The Fe\,{\sc i} lines give the best opportunity for mode identification.
Table~\ref{mtable} lists the 20 strongest unblended lines, where Column~1
is the central wavelength, Column~2 is our line ID number (the first digit
indicates echelle order in which the line falls), Column~3 is the relative
line depth (in percent), Column~4 is the amplitude of the strongest mode
and Column~5 is the uncertainty in this amplitude.  Three of the chosen
Fe\,{\sc i} lines are detected in two echelle orders.  For these lines, the
two measurements of oscillation amplitude agree within the errors.  All the
lines have the same phases within the uncertainties.  As evident from
Table~\ref{mtable}, the amplitude of the main mode changes between the
different lines. This could be due to variations in the temperature
sensitivity of the different Fe{\sc i} lines, or due to blending with other
species with different temperature sensitivities to Fe\,{\sc i}.

\begin{figure}
  \resizebox{\hsize}{!}{\includegraphics[20,0][277,202]{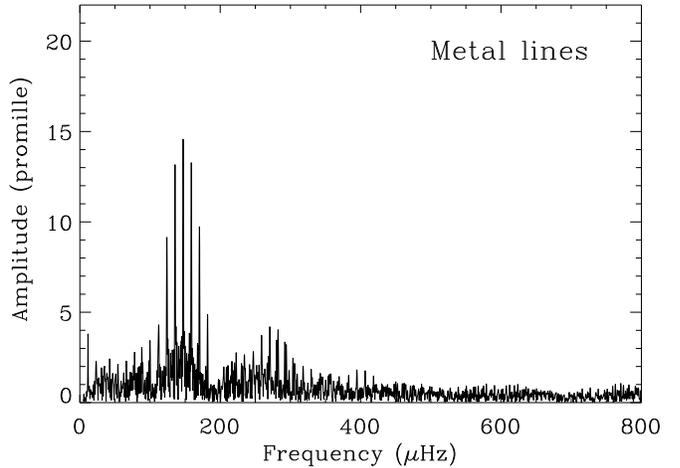}}
  \caption[]{Amplitude spectrum of twenty combined Fe\,{\sc i} lines
  measured through the EW using profile fitting.}  \label{m1}
\end{figure}

Fig.~\ref{m1} shows the amplitude spectrum obtained by combining the EW
time series of the twenty strong Fe\,{\sc i} lines, weighted according to
their signal-to-noise ratios.  To be precise, the combined time series
$w(t)$ was obtained from the 20 individual series $w_i(t)$ using the
following equation
\begin{equation}
   w(t) = \frac{\left(\sum \frac{w_i(t)}{A_i} \mbox{(S/N)}_i^2\right) \;
		\left(\sum A_i \; \mbox{(S/N)}_i^2 \right)} {\left(\sum
		\mbox{(S/N)}_i^2\right)^2},
\end{equation}
where $A_i$ is the amplitude of the 147\,$\mu$Hz mode and $\mbox{S/N}_i =
A_i/\sigma(A_i)$ is its signal-to-noise ratio.  The noise level in
Fig.~\ref{m1} is 0.5 promille in amplitude.  Oscillations are present with
sufficient signal-to-noise in this combined data set to allow the
amplitudes to be used for mode identification.  In the next section we
measure amplitudes for as many modes as possible in each of the observables
(EWs and photometry), to allow ratios to be calculated for mode
identification.

\section{Frequency Analysis} \label{analysis}

Our data set suffers from a poor window function, which produces very high
side-lobes (see Fig.~\ref{ha}), which complicates the determination of
frequencies in the Fourier domain.  
Fortunate\-ly, Breger et
al.~(\cite{bhn95}, \cite{br98}) have determined very precise oscillation
frequencies for FG~Vir from data obtained over many years.  We have
therefore chosen to rely on these frequencies instead of trying to
determine them from our data set.

First, however, we must check whether our data set is consistent with the
main oscillation frequencies found by Breger et al.\ (\cite{bhn95},
\cite{br98}).  By pre-whitening with their frequencies, we found that the
ten strongest pulsation modes of Breger et al.~(\cite{bhn95}) are also
present in our data set.  {}From the residual spectrum it was clear that
more modes are present at lower S/N\@.

Knowing the frequencies in advance reduces the task of determining
amplitudes and phases to a linear problem.  We used the program ISF
(Frandsen et al.~\cite{kapboo}), which is a least-squares routine that
simultaneously fits a series of known frequencies to a given time series.
To maximize the precision, we have fitted all 23 frequencies determined by
Breger et al.\ (\cite{br98}) to each time series, except one frequency
($f_{20}$) which is very close to an alias of one of the other frequencies.
Table~\ref{tspec} lists the amplitudes for \Ha, \Hb\ and the combined
Fe\,{\sc i} spectral lines.  We have used the same mode numbering as Breger
et al.\ (\cite{br98}).  Although all 23 frequencies were fitted, we shall
only use the eight modes for which S/N$>5$ in both \Ha\ EW and photometry
(see also Table~\ref{tphot}).  Fig.~\ref{ha}b shows the residual spectrum
for \Ha\ EW after subtraction of the fit.

\begin{table}
 \caption[ ]{Pulsation modes detected in different spectral lines measured 
            via the EW} \label{tspec}
 \begin{flushleft}
  \begin{tabular}{lcrrrr}  \hline
   ID &\multicolumn{2}{c}{Frequency}&\multicolumn{3}{c}{Amplitudes} \\
      & & &\Ha\ line &\Hb\ line & Fe\,{\sc i} lines \\
   &$\mu$Hz & c/d  & promille & promille & promille \\
   \hline
   $f_{1}$ &147.18 & 12.716 &20.3 &20.2 &14.1 \\
   $f_{2}$ &280.42 & 24.228 & 3.9 & 3.0 & 3.1 \\
   $f_{3}$ &270.87 & 23.403 & 3.9 & 3.4 & 4.0 \\
   $f_{4}$ &243.65 & 21.052 & 3.2 & 3.3 & 1.5 \\
   $f_{5}$ &229.95 & 19.868 & 3.1 & 2.3 & 1.4 \\
   $f_{6}$ &140.67 & 12.154 & 3.7 & 3.6 & 4.0 \\
   $f_{7}$ &111.76 &  9.656 & 4.1 & 4.3 & 1.9 \\
   $f_{8}$ &106.47 &  9.199 & 3.0 & 4.0 & 1.6 \\
   \hline
  \end{tabular}
 \end{flushleft}
\end{table}

\begin{figure}
  \resizebox{\hsize}{!}{\includegraphics[20,-30][277,435]{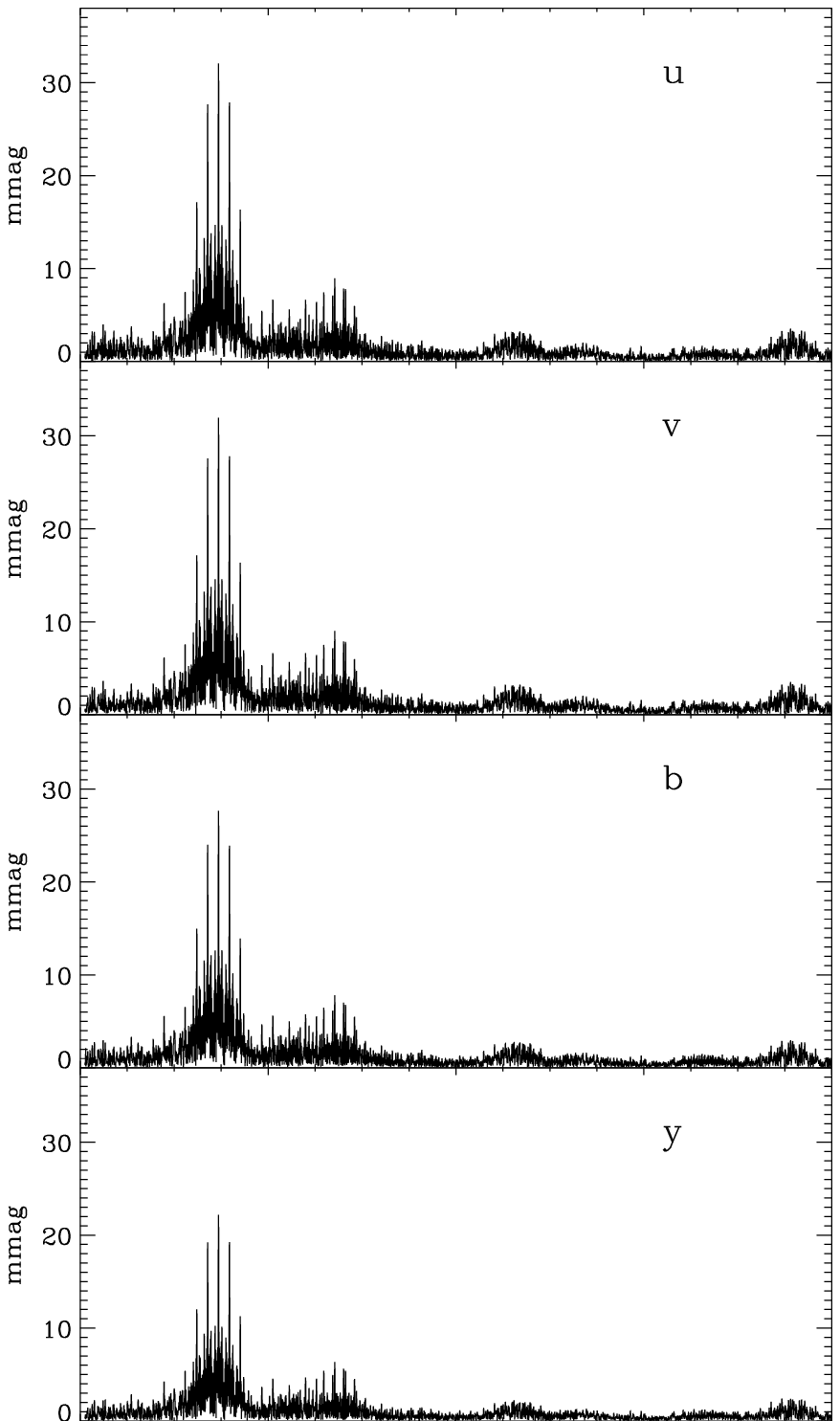}}
  \caption[]{Amplitude spectra in the four Str\"omgren filters}
  \label{col1}
\end{figure}

Fig.~\ref{col1} shows amplitude spectra (in mmag) for the four\linebreak 
\hbox{Str\"omgren} filters.  The signal can clearly be seen in each filter, 
with the expected decrease in amplitude from blue to red.  
Eight frequencies are detected with S/N$>5$, with amplitudes given in 
Table~\ref{tphot}.  The same eight frequencies were detected in the $b-y$ 
index with lower S/N\@.  In the $\beta$ index, only two frequencies could 
be found with adequate S/N but the table includes the amplitudes for all 
eight frequencies for completeness.  The mean noise levels in the residual 
amplitude spectra of the $u$, $v$, $b$, $y$, $b-y$ and $\beta$ time series 
are 0.42, 0.42, 0.40, 0.35, 0.16 and 0.29~mmag, respectively.  These were 
used to calculate the S/N values in Table~\ref{tphot}.

Because the frequencies are known, the amplitudes can be estimated with
better precision than would be expected from the above noise values.  The
formal solution gives standard deviations of the derived amplitudes for the
{\it uvby} data to be 0.07 mmag.  However, from tests made by determining
amplitudes from subsets of the different time series, we find that the main
uncertainty on the amplitudes comes from the very complicated window
function.  Furthermore, some of the modes are closely spaced in frequency
and are not well separated in our data.  This is especially true for the
closest pair of modes $f_2=280.42\mu$Hz and $f_{11}=280.09\mu$Hz, which are
unresolved by our observations.  The derived amplitudes of these modes
change when data are added to or removed from the time series.  In order to
minimize the influence of the window function on the amplitudes, so as to
be able to compare amplitudes from different observables, we have removed a
few nights from the photometric time series to obtain an observing window
similar to that of the La Silla spectroscopic observations.  In this way,
even if amplitudes of some modes are slightly wrong due to the complicated
window function, this effect should be about the same in each time series
and therefore cancel when we calculate the amplitude ratios.

Breger et al.\ (\cite{bhn95}) presented broad-band $V$ photometry for the
ten strongest modes in FG Vir, which we can compare with our Str\"omgren
$y$ amplitudes.  {}For the strongest mode ($f_1$) they obtained an
amplitude of 22.0 mmag, which is in good agreement with our $y$ amplitude.
We also agree on all other amplitudes, within the observing errors, except
the $f_3$ mode at $270\mu$Hz, which has a much higher amplitude in our data
set.  As discussed by Breger et al.\ (\cite{br98}), the amplitude of the
$f_3$ mode has been highly variable.  It has increased from 1.4\,mmag in
1992, 2.3\,mmag in 1993 to 4.1\,mmag in 1995, and our 1996 measurement
gives 5.2\,mmag.

\begin{table*}[ht]
 \caption[ ]{Str\"omgren photometry for FG~Vir} \label{tphot}
 \begin{flushleft}
 \begin{tabular}{lcrrrrrrrrrrrrrrrrr}  \hline
 \noalign{\smallskip}
 ID&Freq.&\multicolumn{15}{c}{Amplitudes}\\
 \cline{3-19}
 \noalign{\smallskip}
 &&\multicolumn{2}{c}{$u$}&\ &\multicolumn{2}{c}{$v$}&\ &\multicolumn{2}{c}{$b$}&\ &\multicolumn{2}{c}{$y$}&\ &\multicolumn{2}{c}{$b-y$}&\ &\multicolumn{2}{c}{$\beta$}\\
 \noalign{\smallskip}
 \cline{3-4}\cline{6-7}\cline{9-10}\cline{12-13}\cline{15-16}\cline{18-19}
 \noalign{\smallskip}
 & $\mu$Hz&mmag &S/N &&mmag &S/N &&mmag &S/N &&mmag &S/N &&mmag &S/N &&mmag &S/N \\ 
 \noalign{\smallskip}
 \hline
 \noalign{\smallskip}
 $f_1$ &147.18 &32.5&77 &&32.4&77 &&27.6&69  &&22.0&63 &&5.4 &33  &&5.3 &18\\
 $f_2$ &280.42 &6.5 &15 &&6.5 &15 &&6.5 &16  &&5.7 &16 &&1.2 &7.3 &&0.7 &2.5\\
 $f_3$ &270.87 &7.1 &17 &&7.3 &17 &&6.5 &16  &&5.2 &15 &&1.0 &6.5 &&1.3 &4.4\\
 $f_4$ &243.65 &4.6 &11 &&4.5 &11 &&4.2 &10  &&3.6 &10 &&0.7 &4.2 &&1.2 &4.0\\
 $f_5$ &229.95 &3.3 &7.8&&3.3 &7.8&&2.8 &7.0 &&2.6 &7.5&&0.4 &2.5 &&0.5 &1.8\\
 $f_6$ &140.67 &7.8 &18 &&7.6 &18 &&6.0 &15  &&5.3 &15 &&0.8 &5.1 &&1.0 &3.4\\
 $f_7$ &111.76 &5.3 &12 &&5.2 &12 &&5.0 &13  &&4.3 &12 &&0.6 &4.0 &&0.9 &3.1\\
 $f_8$ &106.47 &3.9 &9.3&&4.3 &10 &&3.5 &8.8 &&3.0 &8.6&&0.5 &3.3 &&0.8 &2.8\\
 \noalign{\smallskip}
 \hline
 \end{tabular}
 \end{flushleft}
\end{table*}

\section{Mode identification}\label{modeid}

The investigation by Bedding et al.\ (\cite{bkrb96}) into the sensitivity
of various observing methods to modes with different $\ell$ showed that the
Balmer-line EWs have a spatial response similar to that of radial velocity
measurements.  This is because velocity measurements only detect the
component along the line of sight to the star and so, when integrated over
the stellar disk, are preferentially weighted to the central regions.
Balmer lines have a strong centre-to-limb variation, becoming much weaker
towards the limb, and therefore have a response to oscillations similar to
that of radial velocities.  Intensity measurements, on the other hand, give
less spatial resolution because the centre-to-limb variation in intensity
(`limb darkening') is relatively small.  Bedding et al.\ went on to suggest
that this difference in spatial response between simultaneous EW and
intensity measurements could be used to determine the $\ell$ value of a
mode.

They also investigated the temperature dependence of several other
absorption lines (Fe\,{\sc i}, Ca\,{\sc ii}, G-band and Mg~b feature) and found
the Mg~b feature and the Fe\,{\sc i} lines to be especially useful for
oscillation measurements.  These spectral lines have much smaller
centre-to-limb variations than the Balmer lines and so have spatial
responses similar to that of intensity measurements.  This means that mode
identifications can be done with our observations by comparing Balmer EW
amplitudes with either intensity amplitudes or Fe\,{\sc i} EW amplitudes.
Modes with different degree $\ell$ should occupy different locations in an
amplitude-ratio diagram.

A preliminary analysis of our observations was given by Viskum et al.\
(\cite{vdk97}), where only the ratio of Balmer-line EWs to intensity
amplitudes was considered. In that paper, we found a clear grouping of the
different modes in an amplitude-ratio diagram, and proposed a preliminary
mode identification for the ten dominant frequencies.  In this paper, we
present a more thorough investigation of the different measurements,
including the many Fe\,{\sc i} lines present in our spectra.

\begin{figure}
  \resizebox{\hsize}{!}{\includegraphics[20,0][473,456]{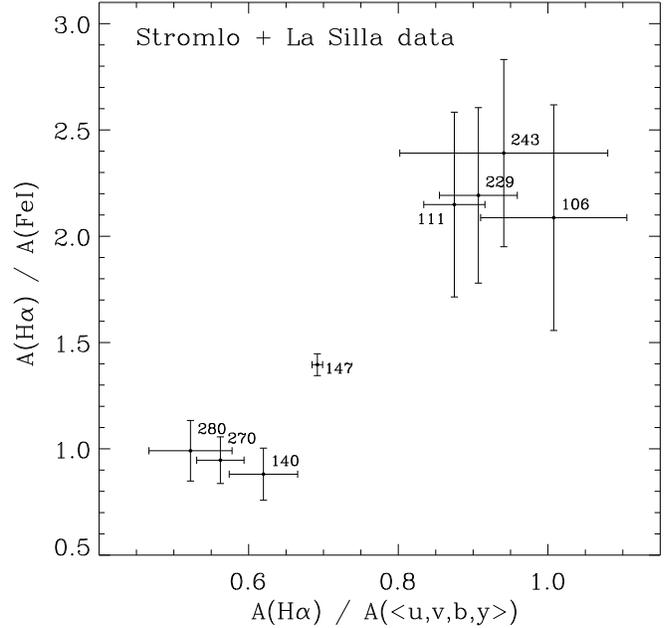}}
  \caption[]{\label{aratio} Amplitude ratios (\Ha/Intensity) versus
  (\Ha/Fe\,{\sc i}) of the eight dominant modes in FG Vir.  The EW
  amplitudes are calculated from the total data set (Mt.\ Stromlo and La
  Silla).  The intensity amplitudes are a mean of the four Str\"omgren
  filters ($uvby$).  Labels show the mode frequencies in $\mu$Hz (truncated
  to integer values).  }
\end{figure}

Fig.~\ref{aratio} shows the amplitude ratios (\Ha/Fe\,{\sc i}) versus
(\Ha/In\-ten\-sity) for the eight dominant modes, obtained using the total \Ha\
data set (Mt.\ Stromlo and La Silla).  Note that amplitudes of oscillations
in EW are measured in promille, while photometric amplitudes are measured
in mmag.  The data points come from Tables~\ref{tspec} and~\ref{tphot},
while the 1-$\sigma$ error bars are based on the tests on data subsets
referred to in Sec.~\ref{analysis}.  The intensity amplitudes are the mean
of the amplitudes in the four Str\"omgren filters ({\it uvby}), in order to
improve the signal-to-noise.  Only the eight dominant modes are shown
because of the large uncertainties on the remaining modes.

The modes are clearly grouped into three regions.  We identify the group at
lower left as radial modes ($\ell=0$), followed by one dipole mode
($\ell=1$) and a group with $\ell \ge 2$.  This trend follows the results
of Bedding et al.~(\cite{bkrb96}), who calculated that the amplitude ratio
between different methods should become greater as $\ell$ increases.  An
important result from this diagram is that the Fe\,{\sc i} lines indeed
have spatial responses similar to that of intensity measurements, as
predicted by Bedding et al.~(\cite{bkrb96}), which shows that mode
identification can be done from spectroscopic measurements alone.  This is
a very important result because it allows one to obtain a mode
identification based on a single data set, with a common window function
and frequency resolution.

\begin{figure}
  \resizebox{\hsize}{!}{\includegraphics[20,0][473,456]{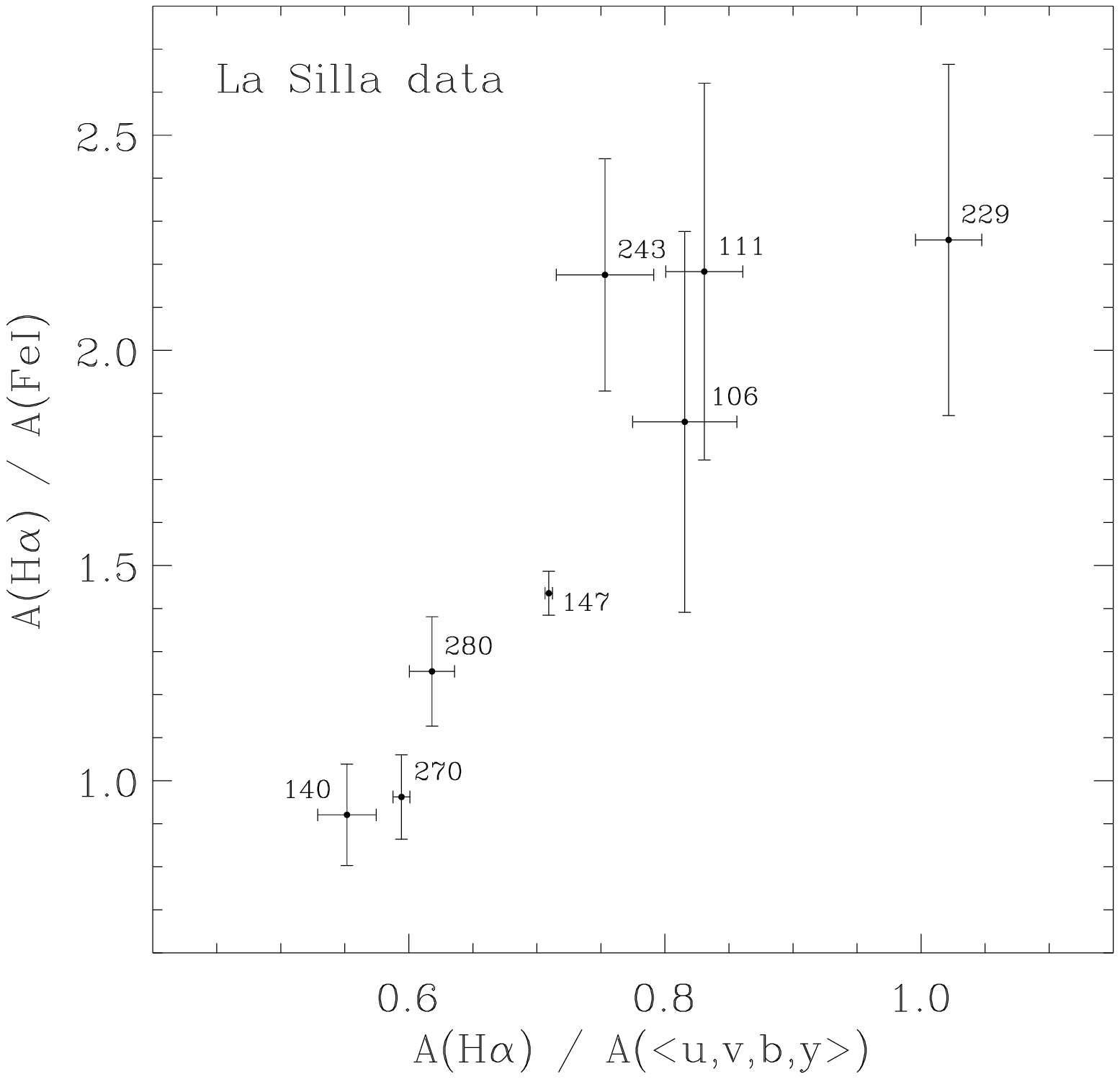}}
  \caption[]{Same as Fig.~\ref{aratio}, but without the Mt.\ Stromlo \Ha\
  data.  } \label{aratio1}
\end{figure}

One problem arises immediately: two of the modes in the lower-left group
($f_2=280.42\,\mu$Hz and $f_3=270.87\,\mu$Hz) are too closely spaced in
frequency for both to be radial modes.  Fig.~\ref{aratio1} shows the same
amplitude ratios, except that the Mt.\ Stromlo \Ha\ data have been
excluded.  The separation of the modes is less clear.  In particular, the
$f_2$ mode has now moved away from the radial group towards the $\ell = 1$
group.  We ascribe this to the change in resolution between the different
data sets, which re-distri\-butes the power between the two unresolved
modes discussed in Sec.~\ref{analysis} ($f_2$ and $f_{11}$).  We have
confirmed using simulations that, because of the presence of $f_{11}$, the
amplitude ratio measured for the $f_2$ mode is very sensitive to the
sampling and to their phase difference.  We expect more reliable results
when the Mt.\ Stromlo data are omitted because the spectroscopic data will
then have a very similar window function to the photometric data.  However,
our simulations show that even the remaining small differences in the
sampling, which inevitably arise from the differences in statistical
weights, will produce systematic errors in the amplitude ratio.  Of the
modes identified by Breger et al.\ (\cite{bhn95}), the pair $f_2$ and
$f_{11}$ have by far the smallest separation (0.32\,$\mu$Hz).  Of those for
which we have EW data, the mode involved in the next closest separation is
$f_4$, which is more than 2\,$\mu$Hz from $f_{14}$.  We therefore do not
expect other modes in our figures to be affected to the same extent as
$f_2$, and we see that they are not.  We conclude that the mostly likely
identification is $\ell=0$ for $f_3$ and $\ell=1$ for $f_2$.

\begin{figure}
  \resizebox{\hsize}{!}{\includegraphics[20,0][473,456]{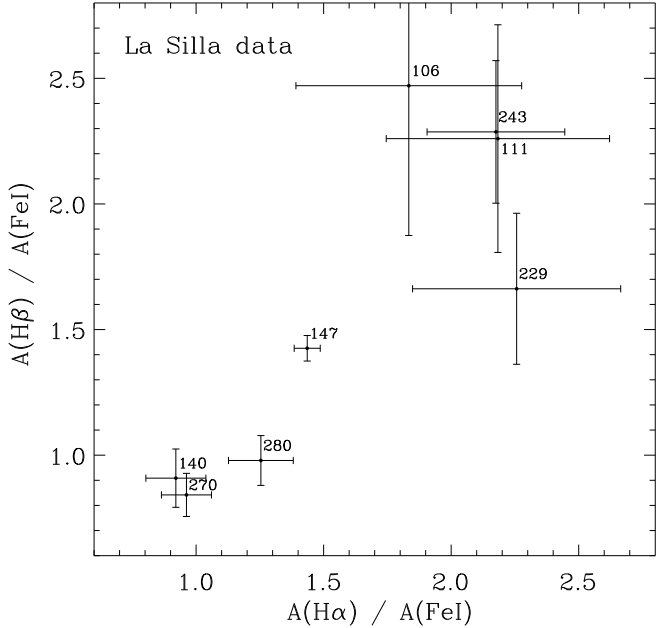}}
  \caption[]{Amplitude ratios (\Hb/Fe\,{\sc i}) versus (\Ha/Fe\,{\sc i}) of
  the eight dominant modes with $1\sigma$ error bars. Only La Silla data}
  \label{aratio2}
\end{figure}

Fig.~\ref{aratio2} shows the amplitude ratios (\Hb/Fe\,{\sc i}) versus 
(\Ha/Fe\,{\sc i}).  The diagram shows that the \Ha\ and the \Hb\ lines have
essentially the same spatial response to the oscillations.  Together, the
three amplitude-ratio diagrams indicate the sensitivity of different
observables applied for mode identifications and the robustness of the
method.

\begin{table}
 \caption[ ]{Mode identifications for FG~Vir } \label{tabmodid}
  \begin{flushleft} \begin{tabular}{lcrcc} \hline ID
  &\multicolumn{2}{c}{Frequency}&\multicolumn{2}{c}{$\ell$-identification}\\
  &$\mu$Hz& c/d & Breger et al.~(\cite{bhn95}) & This work \\ \hline $f_{1}$
  &147.18 & 12.72 & 0 & 1 \\ $f_{2}$ &280.42 & 24.24 & 0 & 1 \\ $f_{3}$
  &270.87 & 23.41 & 2 & 0 \\ $f_{4}$ &243.65 & 21.06 & 1 & 2 \\ $f_{5}$
  &229.95 & 19.87 & 2 & 2 \\ $f_{6}$ &140.67 & 12.16 &\, 2$^1$ & 0 \\
  $f_{7}$ &111.76 & 9.66 &\, 2$^1$ &\, 2$^1$ \\ $f_{8}$ &106.47 & 9.20 &\,
  2$^1$ &\, 2$^1$ \\ \hline \end{tabular} \end{flushleft} $^1$ Possible
  $g$-mode.
\end{table}

\begin{figure}
  \resizebox{\hsize}{!}{\includegraphics[42,20][270,207]{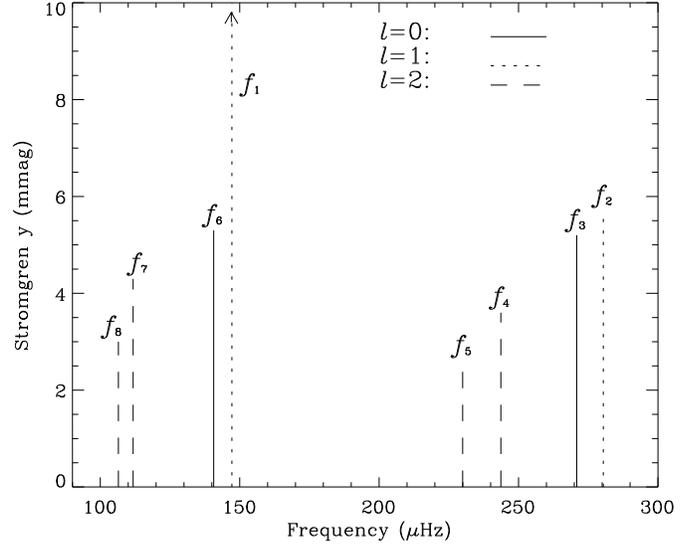}}
  \caption[]{\label{schem} Schematic diagram of the eight strongest modes
  in FG~Vir, with our $\ell$-identifications.  The vertical axis shows the
  photometric amplitude in Str\"omgren~$y$.}
\end{figure}

The mode identifications are summarized in Table~\ref{tabmodid} and
Fig.~\ref{schem}.  In the table we also show the identifications given by
Breger et al.\ (\cite{bhn95}) and as seen the agreement is poor. 
This identification set is actually only one of several possible solutions
 selected by Breger et al.\ (\cite{bhn95}) from a family of
stellar models. They chose this solution, because it
was in agreement with the constraints provided by the spectroscopic 
determination by Mantegazza et al.\ (\cite{mpb94}), who found the $f_1$ mode
to be radial. At the same time Breger et al.\ (\cite{bhn95}) found from
two-colour photometry a negative phase shift 
($\phi_{B-V}-\phi_V=-10.4\pm2^{\circ}$) for the primary frequency ($f_1$). 
This actually excludes the possibility of $f_1$ being a radial mode, and
indeed, this is also the result we obtain from the amplitude ratio diagrams.
Moreover, recent work by Breger et al.\ (\cite{br97comm}) agrees with our
identifications for all eight modes in Table~\ref{tabmodid}.  The earlier
disagreement clearly demonstrates the difficulty of classifying modes by
using comparisons between stellar models and observed frequencies, without 
the help of additional information.

Our identification of the strongest mode as dipole ($\ell =1$) is in good
agreement with this mode's photometric negative phase difference obtained
by Dawson et al.~(\cite{dbl95}).  Another important result is that, as we
show below, the $f_6$ mode is probably the fundamental radial mode ($n=1$,
$\ell=0$)).  In this case, the two modes with lower frequencies ($f_7$ and
$f_8$) must be $g$-modes.  These modes will be particular interesting for
asteroseismology because they contain information about the convective core
and convective overshooting.

The above interpretation is based on the results obtained by Bedding et
al.~(\cite{bkrb96}), who considered the special case in which the rotation
axis points towards the observer, so that contributions only
arise from zonal modes ($m=0$).  Different orientations of the axis and the
influence of non-zero $m$ values on the amplitude-ratio diagrams need to be
investigated in detail.  However, provided the dependence of amplitudes on
$m$ is smaller than the dependence on $\ell$, our identifications are
likely to be correct.

\section{Comparison with a pulsation model} \label{comp}

\subsection{Density of FG Vir}	\label{secdensity}

Once the oscillation modes are identified, the next step in a seismic
analysis is to compare the observed frequencies with models.  Calculating a
full theoretical model for FG Vir is beyond the scope of this paper.
However, much progress can be made by scaling from existing model
calculations.  Here we use calculations by Christensen-Dalsgaard
(\cite{cd93}) for a $2.2M_{\odot}$ star with solar metallicity, generated
during investigations of the $\delta$ Scuti variable $\kappa^2$ Bootis
(Frandsen et al.\ \cite{kapboo}).  Our aim is to make an accurate estimate
of the mean density of FG Vir using the radial modes.  We do this by using
a model that is homologous to FG Vir.

Two stars are homologous if their mass distributions are related by a
simple spatial scaling factor (Kippenhahn \& Weigert \cite{kw}).  That is,
suppose the two stars have masses $M$ and $M'$ and radii $R$ and $R'$, and
the mass enclosed by radius $r$ (or $r'$) is $m$ (or $m'$).  Then at all
points where the relative radii are equal ($r/R = r'/R'$), the relative
enclosed masses will also be equal ($m/M = m'/M'$).  This being the case,
and assuming the stars have the same chemical composition, the oscillation
frequencies of one star can be found by multiplying the frequencies of the
other star by a constant (equal to the square root of the ratio of their
densities).

\begin{figure}
  \resizebox{\hsize}{!}{\includegraphics[140,232][458,546]{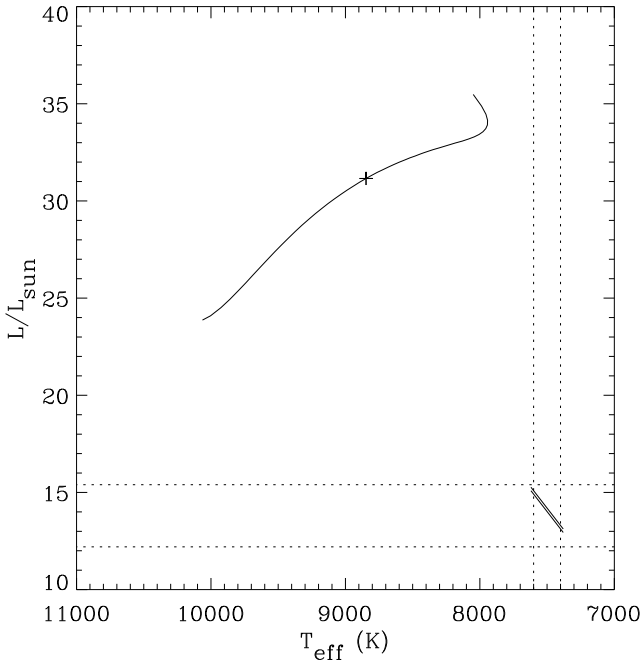}}
  \caption[]{\label{mfig1} H--R diagram showing the evolutionary track of
  the 2.2 $M_{\odot}$ model.  The cross marks the model most closely
  homologous to FG Vir (see Sec.~\ref{secdensity}).  The vertical dotted
  lines show the 1-sigma range in $T_{\rm eff}$ for FG Vir measured by
  Mantegazza et al\ (\cite{mpb94}).  The horizontal dotted lines show the
  1-sigma range in luminosity determined from the Hipparcos distance (see
  Sec.~\ref{secdistance}).  The pair of diagonal lines are contours of
  constant density and define the $\pm 3\sigma$ asteroseismic solution
  given in Table~\ref{models}.  }
\end{figure}

\begin{figure}
  \resizebox{\hsize}{!}{\includegraphics[111,65][524,317]{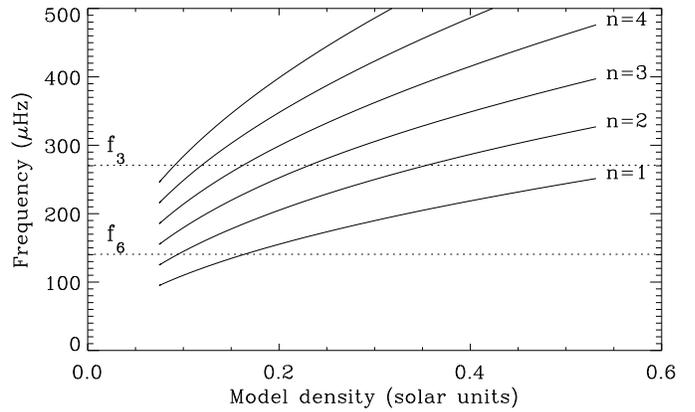}}
  \caption[]{\label{mfigtim1} Theoretical frequencies of the radial
  oscillation modes of a 2.2 $M_{\odot}$ star with solar metallicity
  ($Z=0.02$), plotted as a function of stellar density (the star evolves
  from right to left).  The dotted lines show the frequencies of the two
  observed modes in FG Vir that we have identified as radial.}
\end{figure}

The evolutionary track of the 2.2 $M_{\odot}$ model is shown in
Fig.~\ref{mfig1}.  The model is calculated from the zero-age main sequence
(ZAMS) until the end of the core hydrogen-burning phase.
Fig.~\ref{mfigtim1} shows the frequencies of the radial modes of the
2.2$M_{\odot}$ model, plotted as a function of stellar density.  As the
model star evolves from right to left, the density and the frequencies both
decrease.  We only consider radial modes because those of higher degree
have more complicated time dependencies (Christensen-Dalsgaard \cite{cd93};
Frandsen et al.\ \cite{kapboo}).  The horizontal dotted lines indicate the
two observed modes in FG Vir that we have identified as radial ($f_3$ and
$f_6$).  Since the mass of FG Vir is somewhat less than 2.2 $M_{\odot}$, as
we shall see, our aim is not to find an exact match to the model
frequencies.  Instead, we locate the stage in the evolution of the model at
which the structure of the model is homologous to that of FG Vir.  This
will occur when the model frequencies and those of FG Vir are related by a
single multiplicative factor (which will be the square root of the density
ratio).  The first step is therefore to identify a pair of radial modes in
the model which match the observed frequency ratio ($f_3/f_6 = 1.926$).

\begin{figure}
  \resizebox{\hsize}{!}{\includegraphics[106,65][525,318]{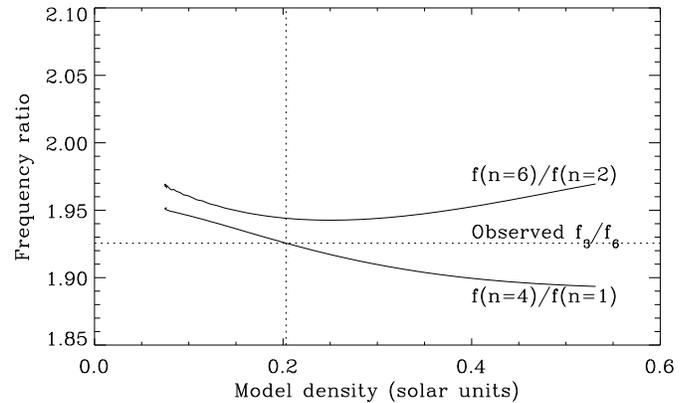}}
  \caption[]{\label{mfigtim2} Frequency ratios for the radial modes of the
  2.2 $M_{\odot}$ model.  The star evolves from right to left.  The
  horizontal dotted line shows the ratio between the two modes in FG Vir
  that we have identified as radial.  The vertical dotted lines shows the
  density of the corresponding model ($0.203\,\rho_{\odot}$).  That model
  should be homologous to FG Vir. }
\end{figure}

Fig.~\ref{mfigtim2} shows frequency ratios for the radial modes of the 2.2
$M_{\odot}$ model.  We see that only one pair of modes matches the observed
frequency ratio, which leads us to identify $f_6$ with $n=1$ and $f_3$ with
$n=4$.  The density of the corresponding 2.2 $M_{\odot}$ model is
$0.203\,\rho_{\odot}$ and we select this as the model most closely
homologous to FG Vir.  To estimate the density of FG Vir, we simply
multiply the density of the homologous model ($0.203\,\rho_{\odot}$) by the
square of the ratio $f_{\rm obs}/f_{\rm model}$.  The result is
$0.164\,\rho_{\odot}$.

We note in passing that the identification of $f_2=280.42\,\mu$Hz as a
radial mode (see Sec.~\ref{modeid}) is inconsistent with the model
frequencies.  It is clear from Fig.~\ref{mfigtim2} that the ratio $f_2/f_6
= 1.993$ does not match the frequency ratio for any of the models.

We can now attempt to identify other radial modes among the published
frequencies.  Based on the above densities, the radial mode with $n=2$
should have a frequency of 185.8\,$\mu$Hz.
Breger et al.\ (\cite{br98}) observed a mode at 186.04 $\mu$Hz ($f_{12}$)
which is a good match.  An exact fit to this observed frequency requires a
density for FG Vir of $0.1647\rho_{\odot}$.  This differs from the above
value by 0.3\%, providing an idea of the uncertainty in this process.  It
seems likely that $f_{12}$ is a radial mode, but its amplitude is too weak
for spectroscopic confirmation.

Using the same argument, the radial mode with $n=3$ should have a frequency
of 228.6\,$\mu$Hz.  
The two observed frequencies nearest this value are $f_5=229.95$ $\mu$Hz
and $f_9=222.55$ $\mu$Hz (Breger et al.\ \cite{br98}).  The best match is
$f_5$ but the discrepancy is 0.6\% (1.2\% in density) and our
amplitude-ratio diagrams indicate that this mode is not radial.  The
frequency $f_9$ differs from the expected value by 2.7\% and we have no
information about its $\ell$ value.  We therefore do not identify either of
these as the radial $n=3$ mode.

\begin{figure}
  \resizebox{\hsize}{!}{\includegraphics[99,65][525,318]{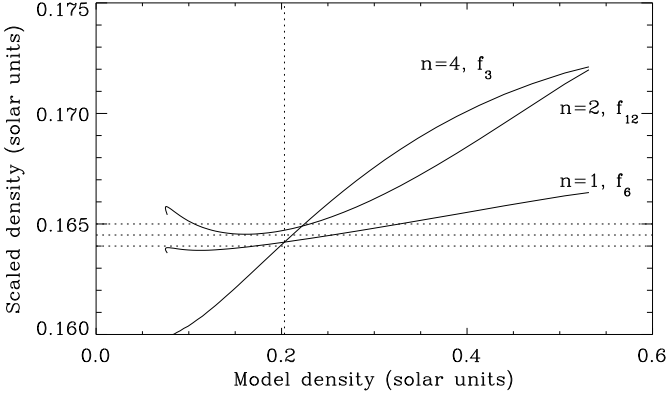}}
  \caption[]{\label{mfigtim3} Scaled density for FG Vir versus model
  density.  For each radial mode, as a function of model density, we plot
  model density multiplied by $(f_{\rm obs}/f_{\rm model})^2$.  The model
  most closely homologous to FG Vir has a density of about
  $0.203\,\rho_{\odot}$ (vertical dotted line).  The derived density for FG
  Vir, shown by the horizontal dotted lines, is $0.1645\pm
  0.0005\,\rho_{\odot}$.}
\end{figure}

To further establish the uncertainty in our density estimate, we have
constructed the plot in Fig.~\ref{mfigtim3} as follows.  For each of the
three observed radial modes, we take each time step in the model
evolution and use the ratio $(f_{\rm obs}/f_{\rm model})^2$ to calculate
the density that FG Vir would have if it were homologous to the model at
that time step.  This scaled density is plotted as a function of the model
density.  We see that the scaled density is fairly insensitive to exactly
which stage in the model evolution is chosen.  In other words, although the
model most closely homologous to FG Vir has a density of
$0.203\,\rho_{\odot}$ (vertical dotted line in Fig.~\ref{mfigtim3}),
adjacent models do quite well (note the scales on the axes differ by more
than a factor of 20).  Indeed, the model with the same actual density as FG
Vir has similar absolute frequencies to FG Vir (to better than 0.5\%),
indicating that even this model is a reasonably good representation of
FG~Vir.

On the basis of this discussion, we can set limits on the density of FG Vir
as shown by the horizontal dotted lines in Fig.~\ref{mfigtim3}, which are
chosen to span the three density estimates.  We then have:
\begin{equation}
 \rho = 0.1645\pm 0.0005\,\rho_{\odot}.	\label{eqrho}
\end{equation}
That is, assuming solar metallicity for FG Vir, we have determined the
density to a precision of about 0.3\%.  It would clearly be desirable to
obtain a spectroscopic measurement of the metallicity of this star, since
this will affect the density at the level of about 1\%.

\subsection{Distance and mass of FG Vir}	\label{secdistance}

The effective temperature of FG Vir was determined spectroscopically by
Mantegazza et al.\ (\cite{mpb94}) to be $7500 \pm 100$\,K.  {}From the
models used in the seismic analysis of $\kappa^2$~Boo (Frandsen et al.\
\cite{kapboo}), we have derived a relationship between luminosity, mass and
effective temperature for stars with this effective temperature.  For solar
metallicity ($Z=0.02$) we find
\begin{equation}
 {L \over L_{\odot}} \simeq 
	0.99 \left ( {M\over M_{\odot}} \right)^{4.8}  
	\left ( {T_{\rm eff}\over 5777\mbox{ K}}\right )^{-0.8}
	\label{eqLMT}
\end{equation} 
Since $L \propto R^2 T_{\rm eff}^4$ and mean density is proportional to
$M/R^3$, we then have:
\begin{equation}
 {\rho \over \rho_{\odot}} \simeq
       1.0021 \left ( {T_{\rm eff}\over 5777\mbox{ K}}\right )
          \left ( {R\over R_{\odot}}\right )^{-2.5833}
	\label{eqTRrho}
\end{equation} 

\begin{table}
 \caption[ ]{Parameters of FG Vir for three different values of temperature} 
   \label{models}
 \begin{flushleft}
  \begin{tabular}{llll}  \hline
  $T_{\rm eff}$ &  7400 K  & 7500 K  & 7600 K      \\
  \hline
  $R/R_{\odot}$ & 2.215 $\pm$ 0.003 & 2.227 $\pm$ 0.003 & 2.238 $\pm$ 0.003 \\
  $L/L_{\odot}$ &13.21  $\pm$ 0.03  &14.08  $\pm$ 0.03  &15.00  $\pm$ 0.03  \\
  $M/M_{\odot}$ & 1.788 $\pm$ 0.008 & 1.816 $\pm$ 0.008 & 1.844 $\pm$ 0.008 \\
  $\log  g^a $   & 3.999 $\pm$ 0.001 & 4.002 $\pm$ 0.001 & 4.004 $\pm$ 0.001 \\
  \hline
  \end{tabular}
 \end{flushleft}
 $^a$ units of $g$ are cm\,s$^{-2}$
 \end{table}

Using the density given in Equation~\ref{eqrho}, we can then estimate the 
radius, luminosity and mass of FG Vir.  Table~\ref{models} shows the results 
for three values of effective temperature.  The quoted uncertainties only
include the uncertainty in the mean density.  The values in
Table~\ref{models} define a locus of constant density in the H--R diagram,
as shown in Fig.~\ref{mfig1}.  The portion of this curve that lies between
the extremes of valid temperatures defines the solution.

\begin{table}
 \caption[]{Parameters of FG Vir based on asteroseismic density$^a$} \label{aprop}
 \begin{flushleft}
 \begin{tabular}{ll}  \hline
     $\rho / \rho_{\odot}$ &  0.1645 $\pm$ 0.0005 \\
     $R/R_{\odot}$         &  2.227  $\pm$ 0.012 \\
     $L/L_{\odot}$         & 14.1    $\pm$ 0.9 \\
     $M/M_{\odot}$         &  1.82   $\pm$ 0.03 \\
     $\log  g $            &  4.002  $\pm$ 0.003 \\
     distance (pc)$^b$     & 84      $\pm$ 3     \\
 \hline
 \end{tabular}
 \end{flushleft}
 $^a$ assuming $T_{\rm eff}=7500\pm 100$ K and $Z= 0.02$. \\
 $^b$ using $L/L_{\odot}$, $V=6.57$, $BC=-0.07$ and $M_{\rm
 bol,\odot} = 4.75$.
\end{table}

The main sources of uncertainty in this process arise from imprecise
knowledge of the temperature and metallicity and from the use of the
relation between $L$, $M$ and $T_{\rm eff}$ (Equations~\ref{eqLMT}
and~\ref{eqTRrho}).  Assuming that the main uncertainty is the temperature,
we find the properties of FG Vir listed in Table~\ref{aprop}.  The errors
are not independent: increasing the temperature will increase the
luminosity, mass and radius.

The derived distance is in excellent agreement with the Hipparcos value of
$83\pm5$\,pc, as is also clear from Fig.~\ref{mfig1}.  It is important to
note that if the temperature had been better known, e.g. to $\pm30$\,K, we
would have obtained an error on the distance of only $\pm 1\mbox{ pc}$
(internal error).  The surface gravity agrees with the value of $\log g =
3.9\pm 0.2$ determined spectroscopically by Mantegazza et al.\
(\cite{mpb94}).  The high precision of our $\log g$ determination arises
because $g$ is given by $\rho R$.

\begin{table}
 \caption[]{\label{hprop} Parameters of FG Vir based on Hipparcos
 distance$^a$}  
 \begin{flushleft}
 \begin{tabular}{ll}  \hline
     distance (pc)         & 83      $\pm$ 5    \\
     $L/L_{\odot}${}$^b$       & 13.8    $\pm$ 1.6  \\
     $R/R_{\odot}$         &  2.20   $\pm$ 0.15 \\
     $M/M_{\odot}$         &  1.82   $\pm$ 0.03 \\
     $\rho / \rho_{\odot}$ &  0.17   $\pm$ 0.03 \\
     $\log  g $            &  4.0    $\pm$ 0.05 \\
 \hline
 \end{tabular}
 \end{flushleft}
 $^a$ assuming $T_{\rm eff}=7500\pm 100$ K and $Z= 0.02$. \\
 $^b$ using distance, $V=6.57$, $BC=-0.07$ and $M_{\rm
 bol,\odot} = 4.75$.
\end{table}

Table~\ref{hprop} shows the fundamental properties for FG~Vir obtained by
combining the Hipparcos distance with Equations~\ref{eqLMT}
and~\ref{eqTRrho}, without using the asteroseismically determined density.
It can be seen that the asteroseismic properties for FG Vir are far more
precise than those obtained by using the Hipparcos data, even with 100\,K
uncertainty in the temperature.  Such calculations show clearly the
potential of asteroseismology.  The present analysis is based on three
$\ell=0$ modes out of more than 20 known frequencies.  We have identified
modes with $\ell=1$ and $\ell=2$ and these modes will contain accurate
information on the detailed evolutionary stage of FG Vir.  This should
provide strong constraints on models computed specifically for FG Vir,
resulting in a fine-tuned model.  The effect from metallicity should also
be investigated.

To check our identification of $\ell =0$ modes, we compared all 24
frequencies published by Breger et al.\ (\cite{br98}) with the frequency
ratios of radial modes in the 2.2M$_{\odot}$ model.  The highest
correlation occurred with the following solution: $f_6$ ($n=1$), $f_{12}$
($n=2$), $f_5$ ($n=3$) and $f_3$ ($n=4$), with a scatter between observed
and best-fit theoretical frequencies of about 1\,$\mu$Hz.  These
identifications are in full agreement with our own results and give us
confidence that we have correctly identified the radial modes in FG Vir.

Finally, we note that the strongest mode ($f_1$) is likely to be $n=1$,
$\ell =1$ and is probably undergoing an avoided crossing with a $g$-mode.
This would make this mode very interesting for testing a more detailed
model.

\section{Rotation period of FG~Vir}

\begin{figure}
  \resizebox{\hsize}{!}{\includegraphics[13,-25][275,341]{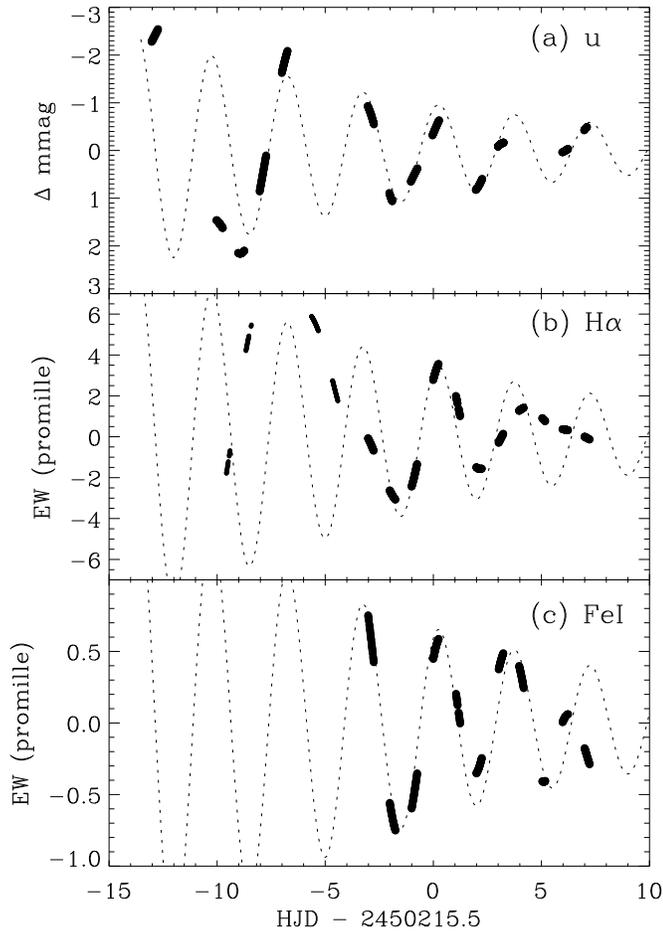}}
  \caption[]{\label{figrot} The raw time series for (a)~the photometric
  light curve in the $u$ filter; (b)~the \Ha\ line EW (small symbols: Mt.\
  Stromlo data, large symbols: La Silla data); (c)~the combined Fe\,{\sc i}
  EW time series.  The dotted curve shows a damped oscillation with a
  period of 3.5 days and with a mean lifetime (the inverse of the damping
  constant) of 14 days.  }
 
\end{figure}

A periodic trend was evident in the raw EW data for all metal lines and in
the two Balmer lines.  The same periodicity was seen in all four
photometric light curves.  The variation is clearly seen after removal of
the known pulsation signal and smoothing (Fig.~\ref{figrot}).

There appears to be a variability with a period of $3.5\pm 0.2$ days and a
decaying amplitude.  Note that the \Ha\ time series is composed of data
from telescopes at two different sites, while the Fe\,{\sc i} and \Ha\ time
series were obtained using different reduction programs.  Furthermore, the
photometric time series were obtained with a third telescope.  All time
series are in phase, indicating that the variation is caused by the same
phenomenon.

We propose that this 3.5-day variation is intrinsic to the star and
reflects the rotation period of FG Vir.  Using the asteroseismic radius in
Table~\ref{aprop}, we then obtain $v_{eq}=33\pm 2$\,km\,s$^{-1}$ for the
equatorial velocity.  Using the value of $v\sin i=21\pm 1$\,km\,s$^{-1}$
found by Mantegazza et al.\ (\cite{mpb94}) gives $i=40\degr\pm 4\degr$ for
the inclination angle, in agreement with the value of $i=31\degr \pm
5\degr$ found by Mantegazza et al.\ from a moments analysis of $f_5$ and
$f_7$.

\section{Summary of results}

We have investigated a new technique to measure the oscillations in
$\delta$~Scuti stars via changes in the equivalent widths of absorption
lines.  An important advantage of this new technique is that only
medium-dispersion spectra are needed, which makes the method suitable for
small and medium-sized telescopes and for multi-site campaigns.  Our main
results are summarized below.
\begin{itemize}

\item Our detection of oscillations in FG~Vir from equivalent-width
    measurements of \Ha, \Hb\ and Fe\,{\sc i} lines is an important
    confirmation of the method developed by Kjeldsen et al.\
    (\cite{kbvf95}) to search for solar-like oscillations.

\item {}From the ratios between oscillation amplitudes measured in EW and
    the four Str\"omgren filters ({\it uvby}), we have identified
    $\ell$-values for eight modes in FG~Vir.

\item We suggest that the two lowest-frequency modes are {\it g}-modes,
    while the strongest mode (147.2$\mu$Hz) is a dipole mode.

\item {}By comparing the frequencies of radial modes with model
     calculations, we obtained a precise density and derived a distance
     that is in excellent agreement with the Hipparcos value.

\item We detected a long-period variation in the time series with a period
    of 3.5 days, which we propose is caused by rotation of the star.

\end{itemize}

\begin{acknowledgements}
We are grateful to M. Breger for providing us with oscillation data of FG
Vir prior to publication and for useful discussions.  We are grateful to
J.V. Clausen for his assistance during the observations with the Danish
0.5m telescope.  We thank F. Kupka and M. Gelbmann for providing us with
spectral line identifications and synthetic spectra for FG Vir, and J.G.
Robertson for many useful comments on the manuscript.  This work was
supported in part by the Danish National Research Foundation through its
establishment of the Theoretical Astrophysics Center, and also by the
Australian Research Council.  We are grateful to the HIPPARCOS group for
making their catalogue available.  This research has made use of the Simbad
database, operated at CDS, Strasbourg, France.
\end{acknowledgements}

\end{document}